\documentclass[a4paper,12pt,epsfig]{iopart}

\begin{document}

\title{Replicating Nanostructures on Silicon by Low Energy Ion Beams}
\author{B. Satpati and B.N. Dev  
\footnote[3]{To whom correspondence should be addressed
(bhupen@iopb.res.in)}}

\address{\ Institute of Physics, Sachivalaya Marg, Bhubaneswar - 751 005,
India}

\begin{abstract} We report on a nanoscale patterning method on Si
substrates using self-assembled metal islands and low-energy ion-beam
irradiation. The Si nanostructures produced on the Si substrate have a
one-to-one correspondence with the self-assembled metal (Ag, Au, Pt)
nanoislands initially grown on the substrate. The surface morphology and
the structure of the irradiated surface were studied by high-resolution
transmission electron microscopy (HRTEM). TEM images of ion-beam
irradiated samples show the formation of sawtooth-like structures on Si.
Removing metal islands and the ion-beam induced amorphous Si by etching,
we obtain a crystalline nanostructure of Si. The smallest structures emit
red light when exposed to a UV light. The size of the nanostructures on Si
is governed by the size of the self-assembled metal nanoparticles grown on
the substrate for this replica nanopatterning. The method can easily be
extended for tuning the size of the Si nanostructures by the proper choice
of the metal nanoparticles and the ion energy in ion-irradiation. It is
suggested that off-normal irradiation can also be used for tuning the size
of the nanostructures. \end{abstract}

\pacs{81.16.-c, 81.07.Bc, 61.80.Jh, 68.37.Lp}

\submitto{\NT}

\maketitle

\section{Introduction} 

The development of large-scale fabrication methods for structured surfaces
with feature sizes in the nanometer range is still a constant challenge in
the nanotechnology community. An accurate control of the surface
topography would have a great impact in the fields of materials science
and nanotechnology. For example, it is well known in the biomedical
materials research that the shape of a surface controls its interaction
with biological components, as has been recently noted \cite{1}. In order
to bring new alternatives to this key issue, numerous strategies, such as
electron-beam lithography, self-assembling methods, soft lithography and
other techniques, have been designed. Each of these methods has advantages
and disadvantages depending on the minimum feature size and the degree of
ordering or the material to be used.

In particular, for the case of nanostructured metal surfaces, novel and
interesting routes have been developed recently by ion sputtering. This
interest arises from the particular modifications of properties of the
nanostructured metal surfaces, such as tribological properties (wear) or
their interaction with electromagnetic radiation, which could be used for
developing diverse technologies \cite{2,3} .

In other instances, a scanning probe like a scanning tunneling microscope
tip \cite{4} or an atomic force microscope cantilever \cite{5} has been
used to engrave features with nanometer resolution. A drawback of scanning
probe nanolithography is the limited scan area, typically, in the
micrometer range, over which the pattern can be engraved. Restructuring
over extended surface areas has been obtained by laser irradiation of
metal and semiconductor surfaces \cite{6}.

For large-scale fabrication of nanostructures on metal surfaces by ion
sputtering, surface erosion has proved to be a very valuable and useful
technique \cite{7}. By this procedure nanodotted patterns and ripple
structures along any desired direction can be achieved both on metals
\cite{8} and semiconductor surfaces \cite{9}. Although the method has
demonstrated itself to be a powerful alternative for producing accurate
nanopatterns, it has the disadvantage of being aggressive (i.e. a single
crystal must be eroded for preparing each nanopatterned surface), making
the method unsuitable for serial fabrication. The crystalline quality of
the nanostructures produced by the sputtering process is doubtful. Thus
the direct use of these nanostructures may not be favourable. However,
this nanostructural pattern may be transfered to produce nanostructures of
other materials \cite{9}.

In order to obtain nanostructures of good crystalline quality
self-assembly would perhaps be the chosen route in many cases. For
example, in spite of the indirect band gap of bulk Ge, Ge islands embedded
in Si layers can be used for light emission \cite{10}. Electron-hole pairs
are captured in Ge islands and recombine by light emission. The necessary
size to achieve confinement effects at room temperature is only a couple
of nanometers. This size is at the limits of conventional lithography. An
alternative is the self-assembly of 3D islands during the heteroepitaxial
growth of Ge on Si. Elastic strain in the growing Ge layer drives the Ge
island growth thereby causing partial strain relaxation. The alloy
material $Si_{x}Ge_{1-x}$ offers the possibility of band-gap engineering.
However, when a $Si_{x}Ge_{1-x}$ thin film is grown on Si, unlike Ge the
island growth does not occur, rather a relatively thicker uniform layer
grows. Dense nanoscale islands of $Si_{x}Ge_{1-x}$ can still be produced
by the method we propose and demonstrate in this paper.

In this report we show that by ion irradiation technique it is possible to
tune the surface morphology on the nanometer scale that is in conformity
with the strategy of using irradiation techniques for producing
large-scale nanopatterned surfaces.

Our scheme of replicating nanostructural patterns on Si substrates is
illustrated in figure 1. First self-assembled metal nanoparticles are
grown on a Si substrate [figure 1(a)] by vacuum deposition of thermally
evaporated metal atoms. This is followed by low energy ion beam
irradiation. When the ions impinge on the empty surface (i.e, between the
interisland space) they amorphize Si upto a depth comparable to the ion
range in Si. However, when the ions impinge on a metal island they pass
through the island and lose energy and consequently cannot penetrate deep
into the underlying Si. As a result the thickness of amorphous Si under
the metal islands is smaller. Thus amorphous Si (and complementarily
crystalline Si) in the near-surface region of the substrate develops a
pattern [figure 1(b)]. The amorphous Si is etched out. The etching process
is carried out in an ultrasonic bath so that the metal islands are ejected
and removed simultaneously leaving a crystalline nanostructured Si [figure
1(c)]. The feature size of the nanostructures can be reduced either by
reducing the size of the metal islands for normal implantation or by
irradiation at an angle with respect to the surface normal while providing
an azimuthal sample rotation around the surface-normal as illustrated in
figure 2.

\section{Experinental procedure} 

Single crystal n-type Si(100) wafers were used for the replica
nanopatterning. Before thin film (nano island film) deposition on Si, the
substrates were cleaned sequentially in ultrasonic baths with methanol,
trichloroethylene, methanol, deionized water and acetone. The native oxide
on Si was not removed. The metal (Ag, Au, Pt) films were prepared by a
conventional vacuum evaporation and an e-beam evaporation technique. An
evaporation chamber was initially evacuated to below $1\times10^{-6}$
mbar. Ag and Au metals (99.99 $\%$ purity) were then vapour-deposited from
a resistively-heated molybdenum boat. The pressure during the vapour
deposition process was kept at $2\times10^{-6}$ mbar, and the thickness of
the deposited Ag and Au was 2 nm as monitored by a quartz crystal
microbalance with a deposition rate of 0.01 nm/s. The e-beam evaporation
of Pt was done at a base pressure of $4\times10^{-7}$ mbar. The thickness
of the deposited Pt was 2 nm. The amount of metal deposited on the Si
surface was also verified by Rutherford backscattering spectrometry
experiments. The thermal deposition of Ag and Au and the e-beam deposition
of Pt were carried out at room temperature of the substrate. Surface free
energies of the metals being much larger than that of the native oxide on
the Si surface, island growth occurs upon metal deposition.

Ion irradiation was carried out with 32 keV Au$^{-}$ ions with a fluence
of $1\times10^{14}$ ions/cm$^2$ at normal ($\sim0^{\circ}$) as well as
off-normal ($\sim60^{\circ}$) angle of incidence. A uniform irradiation
was achived using a 1cm $\times$ 1cm scanned beam. The incident ion
current was kept between 40 nA and 60 nA. Such irradiation results in
penetration of the Au$^{-}$ ions into Si, Ag, Au and Pt film under normal
incidence, up to 23.4, 7.9, 5.6 and 5.0 nm respectively as obtained from
SRIM 2003 range calculation \cite{11}.

Following ion irradiation, the native oxide and the irradiation-induced
amorphous Si were eatched in HF (48\%) in an ultrasonic bath. Transmission
electron microscopy (TEM) measurements were carried out using a JEOL
JEM-2010 (UHR) microscope operating at 200 keV. TEM specimens were
prepared by mechanical thinning followed by ion-beam thinning with a Gatan
precision ion polishing system with an accelerating energy of 3 keV.

\section{Results} 

Ag, Au and Pt films of 2 nm nominal thickness on n-type Si(100) surfaces
have been used for the present study. As the deposited metal grows as
islands, height of each island is much larger than 2 nm. A native oxide of
about 2 nm was present on Si in all the cases prior to thin film
deposition. Figure 3(a) shows a cross sectional transmission electron
microscopy (XTEM) image of an as-deposited Ag film. Ag islands, the Si
substrate and the thin oxide between the metal islands and Si are seen in
figure 3(a). For the as-deposited film, small Ag islands approximately 20
nm in diameter were uniformly observed on the Si(100)  surface. Island
growth is preferred when the interaction among the evaporated atoms
(cohesion) is stronger than the interaction between evaporated atoms and
atoms of the substrate (adhesion) \cite{12}. In the island growth mode the
size of the islands increases as the amount of deposited atoms increases,
which has been observed and reported previously \cite{13}. Figure 3(b)
shows an XTEM image of Ag/Si thin film [similar sample as in figure 3(a)]
irradiated with a fluence of $1\times 10^{14}$ ions/cm$^2$ at an impact
angle of $\sim0^{\circ}$ (normal incidence). Below the islands we observe
dark irregular saw-tooth-like dark boundaries bounding crystalline Si. The
lighter region below the islands and the native oxide is amorphous Si. The
surface of the crystalline Si is patterned from the shadowing due to the
metal islands. As the ions penetrate they have a lateral displacement
component due to transverse straggling \cite{11}. The shadowing is
somewhat blurred because of transverse straggling of ions. However, the
nanocrystalline features on the Si substrate have a one-to-one
correspondence with the metal islands.

Figure 4(a) shows an XTEM image of a Ag/Si thin film irradiated with a
fluence of $1\times 10^{14}$ ions/cm$^2$ at an impact angle of
$\sim60^{\circ}$. In this tilted angle geometry, the hillock patterns of
crystalline Si are not directly under the islands. Figure 4(b) shows an
XTEM image of an irradiated film etched in HF for 20 min. From figure 4(b)
it appears that the amorphous Si has been etched in HF and Ag islands have
fallen down into the vallies between nano crystalline hillocks. The gap
between the islands and the crystalline Si indicates that some oxide is
still present and longer etching is required for complete removal of
oxide. In addition, to remove these islands from the surface we performed
the HF etching in an ultrasonic bath for 30 min. The XTEM image of this
sample is shown in figure 4(c). Here the Ag islands are absent and only
nanocrystalline Si hillocks are present on the surface. Figure 4(d) is the
magnified portion of the Si substrate marked by a rectangle in figure 4(c)
showning a high resolution (HR) XTEM image. The lattice resolution image
in figure 4(d) shows the boundaries of nanocrystalline hillocks.

Figure 5(a) shows an XTEM image of an as-deposited Au film. Here the
particle size is smaller ($\sim$ 10 nm) than in the case of Ag. Figure
5(b) and 5(c) show XTEM images of Au/Si thin films irradiated with a
fluence of $1\times 10^{14}$ ions/cm$^2$ at impact angles of
$\sim0^{\circ}$ and $\sim60^{\circ}$ respectively. Figure 5(d) is a
HR-XTEM image of the magnified portion of the Si substrate marked by a
rectangle in figure 5(b). A prominant nanocrystalline Si hillock is
observed in this lattice image. In figures 5(b) and 5(c), we also notice
the embedding of Au islands into Si. We observed similar features in MeV
ion irradiation, where embedding of Au islands was more prominent compared
to Ag islands. Additionally, embedded Au islands reacted with Si to form
gold-silicide \cite{14}. The embedding was also more prominent at
$60^{\circ}$ ion impact angle compared to $0^{\circ}$ as observed here. As
the ions penetrate a smaller depth for $60^{\circ}$-impact compared to
$0^{\circ}$-impact, the amorphous Si layer is thinner in figure 5(c)
compared to that in figure 5(b).

Looking at figure 3(a) and figure 5(a), where the island heights are
typically $\sim$ 20 nm and $\sim$ 10 nm, respectively, the island density
appears to be too high to be consistent with a nominal thickness of 2 nm.
However, there is no inconsistency. When looked from the top [see figure
7(a)], the islands actually cover a small fraction ($\sim$ 25\%) of the
surface. In the cross-sectional view, such as in figure 3(a) and figure
5(a), islands at different lateral positions along the depth (i.e.,
perpendicular to the plane of the figure) are also visible. That is why
island density appears to be high in the XTEM images such as figure 3(a)
and 5(a).  This is also the reason why overlapping islands with varying
contrast are seen. Actually a Si hillok is formed under each metal island
seen in the plan-view image like figure 7(a).

For normal incidence irradiation, crystalline Si hillocks are formed
directly under the metal nanoparticles, as seen in figure 3(b). For tilted
ion incidence the Si crystalline hillocks are laterally displaced compared
to the metal particles that produce these Si hillocks by shadowing, as in
figure 4(a). This is expected as illustrated in figure 2(b). In order to
produce smaller Si hillocks, by tilted incidence, according to the scheme
proposed in figure 2, the substrate has to be rotated around the surface
normal during irradiation. We could not try this method as we do not have
the facility for substrate rotation in our irradiation chamber. However,
as we proposed in the last paragraph of section 1, there is another way to
produce nanostructures of smaller sizes. That is, by making the deposited
metal particles smaller. This can, in principle, be done by choosing a
metal of higher surface free energy so that the surface free energy
difference between the metal and the substrate is larger. We choose Pt for
this purpose and the results are discussed below. (Earlier we have used
this concept of surface free energy difference to grow Ge nanoparticles on
polymer-coated Si substrates \cite{15}).

Figure 6(a) shows an XTEM image of an as-deposited Pt film. Here the
particle size is much smaller ($\sim$ 2 nm) than in case of Ag and Au.
Here the lattice image of crystalline Si and the native oxide layer
between the islands and crystalline Si are clearly seen. Figure 6(b) shows
a HR-XTEM image of the film shown in 6(a) following ion irradiation with a
fluence of $1\times 10^{14}$ ions/cm$^2$ at an impact angle of
$\sim0^{\circ}$ and HF etching in ultrasonic bath for 30 min. A
nanocrystalline Si hillock of $\sim$ 6 nm diameter is seen in the lattice
image of figure 6(b). This sample in figure 6(b) shows a reddish tinge
when exposed to ultraviolet light of 275 nm wavelength. It is known that
Si nanoparticles can emit red, green, blue light emission depending on the
particle size \cite{16}. A 6 nm diameter Si particle is expected to emit
red light \cite{16}.

For Ag, Au and Pt nanoparticles grown on the native oxide layer atop the
Si substrate, we notice a general trend $\full$ Ag islands are the largest
and the Pt islands are the smallest. This may be understood from their
respective surface free energies. The surface free energies of Ag, Au and
Pt are 1.25 J/m$^2$ \cite{17}), 1.55 J/m$^2$ \cite{17}) and 2.55 J/m$^2$
\cite{17} respectively. Consequently the largest difference in surface
free energy between Pt and the native oxide (0.3 J/m$^2$ \cite{18}) on
which it is growing is apparently responsible for the growth of smallest
islands in the case of Pt. For a given metal, the size of the islands can
be controlled by controlling the substrate temperature during deposition
\cite{19}.

Spherical monodisperse nanoparticles of much smaller average size,
compared to what is shown here, can be prepared by ion beam sputtering of
samples like those shown in figure 3(a) and 5(a). Upon irradiation,
nanoparticles are ejected from those on the substrate \cite{19,20}. When
these sputtered particles are captured on a Si wafer and this Si wafer
with smaller nanoparticles is used for replicating nanostructures by the
method described here, much smaller nanostructural patterns can be created
on Si. We have prepared Au nanoparticles of an average diameter of $\sim$
5 nm on catcher Si wafers during ion sputtering of samples such as that
shown in figure 5(a). An example of this is shown in figure 7. The ejected
Au particles are much smaller than the size of the original deposited Au
particles. More details of this aspect will be presented elsewhere
\cite{21}.

A way to obtain further smaller Si nanostructures from, say, those in
figure 6(b) is to let it be oxidized. Assuming the same oxidation
behaviour of Si wafers and Si nanocrystals, $\sim$ 2 nm thick native oxide
will form along the outer boundary of the nanocrystal. This would reduce
the diameter of the Si nanocrystalline features from $\sim$ 6 nm to $\sim$
2 nm, which is expected to emit light of shorter wavelength $\full$ close
to blue light \cite{16}.

Pyramid- and dome-shaped self-assembled quantum dots of Ge grows on Si
following the Stranski-Krastanov growth mode \cite{22}. Ge islands grow on
three atomic layers of uniform Ge. However, $Si_{x}Ge_{1-x}$ layers grow
on Si uniformly to much larger thicknesses depending on the composition
(the value of $x$). $Si_{x}Ge_{1-x}$ nanostructures can be produced from
this uniform $Si_{x}Ge_{1-x}$ layer. In order to obtain $Si_{x}Ge_{1-x}$
nanoislands on Si, the method presented here can be conveniently used
provided an appropriate etchant is found for amorphous $Si_{x}Ge_{1-x}$.
This would provide a wider tunability of optical devices using
$Si_{x}Ge_{1-x}$ as the band gap of $Si_{x}Ge_{1-x}$ is tunable over a
wide energy range.

For optoelectronic applications the quantum yield of the Si nanostructures
is of great importance. The aspect of quantum efficiency, which has not
been addressed here, is to be explored in future work.

\section{Conclusions}

Here we have proposed and demonstrated a method of replicating
nanostructures on Si surfaces using metal nanoparticles on Si as mask and
low energy ion-irradiation. Although the nanostructures on Si are not
exact replicas of the overlying metal islands there is a one-to-one
correspondence. When the metal islands are small enough the features of
nanocrystalline Si replicated on the Si surface are also small - small
enough to show optical emission within the visible band. With a proper
choice of etchant the method can be used for a larger number of substrates
besides Si. Although HF does not etch crystalline Si, here we have found
that amorphous Si can be effectively etched in HF.

\section{Acknowledgments}

The authors would like to thank B. Joseph for help in low energy ion
irradiation and D. Kabiraj for providing Pt films on Si. Helpful
discussion with Dr. P.V. Satyam is gratefully acknowledged.

\Figures

\begin{figure} \caption{\label{label}Scheme showing the different steps
involved in the replica nanopatterning method with the ion beam incident
along the surface-normal direction. (a) Metal nanoparticles on Si, (b) ion
irradiation and creation of patterned amorphous Si, (c) after etching out
amorphous Si, nanostructures of Si are revealed.} \end{figure}

\begin{figure} \caption{\label{label}Illustration of how smaller
nanostructures on Si can be produced by tilted implantation and azimuthal
rotation of the substrate. (a) Metal nanoparticles on Si, (b) normal
irradiation (left), off-normal irradiation with azimuthal rotation
(right), (c) larger and smaller Si nanostructures corresponding to the
cases in (b).} \end{figure}

\begin{figure} \caption{\label{label}XTEM images: (a) As-deposited Ag
nanoisland film; a thin ($\sim$ 2 nm) native oxide layer between Ag
islands and crystalline Si is seen, (b) Ag film in (a) irradiated with 32
keV Au$^-$ ion at a fluence of $1\times 10^{14}$ ions /cm$^2$ with the ion
beam incident along the surface normal ($0^{\circ}$ impact angle). The
structures formed underneath the Ag islands have a one-to-one
correspondence. An amorphous Si layer ($\sim$ 20 nm thick) is formed
between the Ag islands and the underlying crystalline Si.} \end{figure}

\begin{figure} \caption{\label{label}XTEM images: (a) The film in 3(a)
irradiated with 32 keV Au$^{-}$ ions at a fluence of $1\times 10^{14}$
ions /cm$^2$ at $60^{\circ}$ impact angle, (b) the irradiated sample in
(a) etched in HF for 20 min; the amorphous Si layer between Ag islands and
the crystalline Si is now thinner, (c) the sample in (a) etched in HF for
30 min in an ultrasonic bath; the islands, the native oxide, and amorphous
Si are removed, (d) the lattice image of a magnified portion of the image
shown in (c) marked with a rectangle.} \end{figure}

\begin{figure} \caption{\label{label}XTEM images: (a) An as-deposited Au
nanoisland film, (b) the Au film in (a) irradiated with 32 keV Au$^-$ ions
at a fluence of $1\times 10^{14}$ ions /cm$^2$ at $0^{\circ}$ impact
angle, (c) the Au film in (a) irradiated with 32 keV Au$^-$ ions at a
fluence of $1\times 10^{14}$ ions /cm$^2$ at $60^{\circ}$ impact angle,
(d) the lattice image of a magnified portion of the image shown in (b)
marked with a rectangle.} \end{figure}

\begin{figure} \caption{\label{label}XTEM images: (a) An as-deposited Pt
nanoisland film; below the Pt islands the sharp native-oxide/Si interface
is seen, (b) the Pt film in (a) irradiated with 32 keV Au$^-$ ion at a
fluence of $1\times 10^{14}$ ions /cm$^2$ at $0^{\circ}$ impact angle and
then etched in HF for 30 min. in ultrasonic bath. The lattice image of a
Si island ($\sim$ 8 nm height and $\sim$ 7 nm diameter) is seen. The sharp
flat Si interface in (a) is nanostructured as shown in (b) by the ion-beam
replica nanopatterning method.} \end{figure}

\begin{figure} \caption{\label{label}Plan-view TEM images of an (a)  
as-deposited Au nanoisland film (nominal thickness 1.3 nm) on Si and (b)  
of ejected Au islands collected on a catcher Si wafer during ion
irradiation (1.5 MeV Au$^{2+}$ ions, fluence : 1$\times$ 10$^{14}$
ions/cm$^2$) of the sample in (a). (c) and (d) show island (lateral) size
distributions for the cases in (a) and (b) respectively. The most probable
size (diameter)  in (c) is $\sim$ 11.5 nm (Gaussian fit) and that in (d)
is $\sim$ 4.5 nm (log-normal fit).} \end{figure}

\end{document}